\documentclass[preprint,5p,twocolumn,10pt]{elsarticle}

\usepackage[pdftex,hypertexnames=false,linktocpage=true]{hyperref}
\hypersetup{colorlinks=true,linkcolor=red,anchorcolor=darkgreen,citecolor=green!50!blue,filecolor=darkgreen,urlcolor=green,bookmarksnumbered=true,pdfview=FitB}

\usepackage{lipsum}
\usepackage{mathtools,cuted}

\usepackage{amsmath}

\usepackage{graphicx}

\usepackage{array}
\usepackage{booktabs}
\usepackage{arydshln}
\usepackage{multirow}

\newcommand{\ie}{\textit{i.e.}}

\usepackage{amsmath,amssymb,mathtools} 
\ifpdftex
    \usepackage{bbm}

\else
    \usepackage[bold-style=ISO]{unicode-math}
\fi

\usepackage{siunitx}  
\makeatletter
\newcommand{\vast}[1]{\bBigg@{#1}}
\makeatother

\newcommand{\intd}[3][]{\ifthenelse{\isempty{#3}}{\mathrm{d}^{#1} #2}{\frac{\mathrm{d}^{#1} #2}{#3}}\;}

\usepackage{stackengine}

\newcommand{\cond}{\; | \;}
\newcommand{\setcond}[2]{\ifthenelse{\isempty{#2}}{\{#1\}}{\{#1\cond{}#2\}}}

\usepackage{braket}
\usepackage{cancel}
    
\usepackage{braket}

\begin{document}

\begin{frontmatter}
\title{Transverse-momentum-dependent gluon distributions of proton within basis light-front quantization}

\author[ynu]{Hongyao Yu\corref{cor1}}
\ead{yuhongyao@stu.ynu.edu.cn}

\author[kek]{Zhi Hu\corref{cor1}}
\ead{huzhi0826@gmail.com}

\author[imp,ucas,keylab]{Siqi Xu\corref{cor1}}
\ead{xsq234@impcas.ac.cn}

\author[imp,ucas,keylab]{Chandan Mondal\corref{cor1}}
\ead{mondal@impcas.ac.cn}

\author[imp,ucas,keylab]{Xingbo Zhao\corref{cor1}}
\ead{xbzhao@impcas.ac.cn}

\author[iowa]{James P. Vary\corref{cor1}}
\ead{jvary@iastate.edu}

\author[]{\\\vspace{0.2cm}(BLFQ Collaboration)}

\address[ynu]{School of Physics and Astronomy, Yunnan University, Yunnan, 650000,China}
\address[kek]{High Energy Accelerator Research Organization (KEK), Ibaraki 305-0801, Japan}
\address[imp]{Institute of Modern Physics, Chinese Academy of Sciences, Lanzhou, Gansu, 730000, China}
\address[ucas]{School of Nuclear Physics, University of Chinese Academy of Sciences, Beijing, 100049, China}
\address[keylab]{CAS Key Laboratory of High Precision Nuclear Spectroscopy, Institute of Modern Physics, Chinese Academy of Sciences, Lanzhou 730000, China}
\address[iowa]{Department of Physics and Astronomy, Iowa State University, Ames, IA 50011, USA}

\cortext[cor1]{Corresponding author}

\begin{abstract}
  Gluon transverse-momentum dependent distributions (TMDs) are very important for revealing the internal structure of the proton. They correspond to a variety of experiments and are among the major tasks of the future electron-ion colliders. In this paper, we calculate the T-even gluon TMDs at the leading twist within the Basis Light-front Quantization framework. We employ the light-front wave functions of the proton obtained from a light-front quantized Hamiltonian with Quantum Chromodynamics input determined for its valence Fock sector with three constituent quarks and an additional Fock sector, encompassing three quarks and a dynamical gluon, with a three-dimensional confinement. After obtaining the numerical results we investigate the properties of gluon TMDs by checking whether they satisfy the Mulders-Rodrigues inequalities and investigate the correlations between transverse and longitudinal degrees of freedom. With their connection to unintegrated gluon distributions in mind, we further investigate the limiting behaviours of gluon TMDs in the small-$x$ and large-$x$ regions.
\end{abstract}
\end{frontmatter}

\section{Introduction}
The internal structure of the proton has been one of the central topics for high-energy physics for decades. Accumulated evidence from past experimental and theoretical investigations show that quarks only contribute a small fraction of the proton spin and mass \cite{protonspin1,SpinMuon:1998eqa,E155:2002iec,Alexandrou:2020sml,Jaffe:1989jz,Ji:1996ek,Ji:2021mtz,Ji:2020baz}. On the theory side, protons are QCD bound states with both quarks and gluons as explicit degree of freedom. Thus, it is very important to investigate the gluon contents inside the proton.

With the help of factorization theorems, various distribution functions have become strong bridges between purely theoretical constructions from the quantum fields and the cross sections of high-energy experiments \cite{DeRoeck:2011na,Bacchetta:2006tn,Diehl:2003ny,Diehl:2011yj,collins_2011,Collins:1981uk,Ji:2004xq,Ji:2005nu}. As the only possible means of peering into the internal structures of the nucleon, the current investigations of the distribution functions have evolved from the simple one dimensional picture of parton distribution functions (PDFs) and form factors (FFs), to detailed three dimensional pictures of transverse-momentum dependent distributions (TMDs) and generalized parton distributions (GPDs) \cite{Constantinou:2020hdm,Boer:2011fh,Boer:2015vso}.

In this paper, we focus on the gluon TMDs, which encode the correlations between parton transverse momenta and the polarizations of both partons and nucleons. The definitions and theoretical investigations of the gluon TMDs date back to the beginning of this century \cite{Mulders:2000sh,Lorce:2013pza,Boer:2016xqr,Kumano:2019igu}. Beginning around 2010, many studies also reveal the non-trivial connections between gluon TMDs and the so-called unintegrated gluon distributions commonly encountered in small-$x$ physics \cite{Dominguez:2011wm,Dominguez:2010xd,Petreska:2018cbf,Xiao:2017yya,Fujii:2020bkl,Altinoluk:2020qet}. Contrary to their importance, currently there are few theoretical calculations available for the numerical values of gluon TMDs \cite{Bacchetta2020,Chakrabarti_2023,Bacchetta:2021twk,more_in_TMD,PhysRevD.94.094022,Bacchetta:2021lvw,Sumrule,PhysRevD.104.014001,YAO2019361}. Here, we work within Basis Light-front Quantization (BLFQ) \cite{initial_BLFQ} and extend the Fock-sector truncation to include one dynamical gluon in the proton to numerically compute the gluon TMDs. 

This paper is organized as follow. We introduce the BLFQ framework in Sec. 2. The formulas of the gluon TMDs expressed by overlapping of the second Fock-sector light-front wave functions are discussed in Sec. 3 along with a set of calculated results. We investigate additional detailed properties of the gluon TMDs in Sec. 4 and conclude in Sec. 5.

\section{BLFQ calculation of the proton with one dynamical gluon\label{Sec2}}
Within the BLFQ framework we obtain light-front wave functions (LFWFs) by solving the following eigenvalue equation:
\begin{align}
\label{1}
    P^-P^+|P,\Lambda\rangle=M^2|P,\Lambda \rangle,
\end{align}
where $|P,\Lambda\rangle$ is the bound state with momentum $P=~(P^+=\sum\limits_ip_i^+,P^-=\frac{M^2}{P^+},P^\perp=0^\perp)$, $P^+$ and $P^-$ represent the longitudinal momentum of the system and the light-front Hamiltonian respectively, $p_i^+$ denote the longitudinal momentum of the $i$-th parton, and $\Lambda$ is the light-front helicity  \cite{LF-helicity,LFQFT_Brodsky_1998}. $M^2$ is the square of the proton mass. We then expand the bound state of the proton in Fock-space and truncate at the Fock-sector with one dynamical gluon which can be schematically expressed as \cite{Siqi2022qqqg}
\begin{align}
    |P,\Lambda\rangle=\Psi^\Lambda_{\mathcal{N}=3}|qqq\rangle+\Psi^\Lambda_{\mathcal{N}=4}|qqqg\rangle,
    \label{focksector}
\end{align}
where $\Psi^\Lambda_{\mathcal{N}=3}$ and $\Psi^\Lambda_{\mathcal{N}=4}$ represent LFWFs corresponding to the three-particle and four-particle Fock-sectors respectively. The $|P,\Lambda\rangle$ introduced in Eq.~(\ref{1}) represents the full eigenstate which is a sum over all potential sets $\{p_i, \lambda_i\}$. 

With the truncation in Eq.~(\ref{focksector}), in this work, we have both gluon and quark as explicit degrees of freedom and adopt the following effective light-front Hamiltonian $P^-=P^-_{\textrm{QCD}}+P^-_\textrm{C}$. $P^-_{\textrm{QCD}}$ is the light-front QCD Hamiltonian restricted to the  $|qqq\rangle$ and $|qqqg\rangle$ Fock-sectors \cite{Siqi2022qqqg}
\begin{align}
    P^-_{\textrm{QCD}}&=\int d^2x^\perp dx^- \left\{ \frac{1}{2}\Bar{\psi}\gamma^+\frac{m^2_0+(i\partial^\perp)^2}{i\partial^+}\psi \nonumber\right.\\
    &+\frac{1}{2}A^i_a[m^2_g+(i\partial^\perp)^2]A^i_a+g_s\Bar{\psi}\gamma_\mu T^a A^\mu_a\psi \nonumber\\
    &\left.+\frac{1}{2}g^2_s\Bar{\psi}\gamma^+T^a\psi\frac{1}{(i\partial^+)^2}\Bar{\psi}\gamma^+T^a\psi  \right\} . 
    \label{P^-_QCD}
\end{align}
The first two terms in Eq.~(\ref{P^-_QCD}) are the kinetic energies of quark and gluon with bare mass $m_0$ and $m_g$, where $\psi$ and $A^i_a$ represent quark and gluon fields, respectively. $\gamma^\mu$ is the Dirac matrix, and $T$ is the generator of $SU(3)$ gauge group in the color space. Although gluon mass in QCD is zero, we admit a phenomenological gluon mass in our low-energy model \cite{Siqi2022qqqg} which, among other advantages, facilitates a description of the nucleon FFs. The third and fourth terms are the quark gluon vertex and instantaneous gluon interactions, respectively, with coupling constant $g_s$. Following the spirit of Fock-sector dependent renormalization \cite{Perry:1990mz,Karmanov:2008br,FSDR_Kamanov}, we introduce a mass counter term ($\delta m_q$) so that the renormalized quark mass is $m_q=m_0-\delta m$ in our leading Fock-sector. We also permit an independent quark mass $m_f$ in the vertex interaction \cite{quark_mass_mf}. 

We introduce confinement in the $|qqq\rangle$ Fock-sector as:
\begin{align}
    P^-_CP^+=\frac{\kappa^4}{2}\sum\limits_{i\neq j}\left\{\Vec{r}_{ij\perp}^2-\frac{\partial_{x_i}(x_ix_j\partial_{x_j})}{(m_i+m_j)^2}\right\}. 
\end{align}
$\Vec{r}_{ij\perp}=\sqrt{x_ix_j}(\Vec{r}_{i\perp}-\Vec{r}_{j\perp})$ is the holographic variable \cite{HQCD_confinement_Brodsky}, $\partial_x\equiv(\partial/\partial x)_{r_{ij\perp}}$, and $\kappa$ is the strength of the confinement. In the $|qqqg\rangle$ Fock-sector, we omit the explicit confinement and hope that the massive gluon and limited basis in our approach sufficiently account for confinement. 

For our BLFQ basis space, we adopt a plane-wave basis in the longitudinal direction, which is confined in a one-dimensional box with length $2L$ and accompanied by antiperiodic (periodic) boundary condition for the fermion (boson). Therefore, the longitude momentum is $p^+=2\pi k/L$, where $k$ is our longitudinal quantum number with half-integer (integer) value for the fermion (boson).  In the transverse plane, we adopt the two-dimensional harmonic oscillator (2D-HO) basis functions $\Phi_{nm}(\Vec{p}_\perp;b)$ with energy scale $b$ \cite{2D-HO_by_Li}. $n$ and $m$ are the principle and orbital momentum quantum numbers for 2D-HO, respectively. We also introduce $\lambda$ to  represent the light-front helicity. In conclusion, for each Fock-particle basis state, there are four quantum numbers $|\alpha_i\rangle=|k_i,n_i,m_i,\lambda_i\rangle$ and the many-body basis states are identified as the direct product of Fock-particle basis states $|\alpha\rangle=\otimes_i |\alpha_i\rangle$. We work in a basis of total color singlet states.  We note that $|qqqg\rangle$ has two color singlet states, so we need an additional index to identify the specific color singlet state. 

We truncate the infinite basis by introducing cutoffs $K$ and $N_{\mathrm{max}}$ in the longitudinal and transverse direction, respectively. The dimensionless variable $K=\sum\limits_{i}k_i$, represents the total longitudinal momentum of the proton and thus the longitudinal momentum fraction of a parton can be expressed as $x_i=k_i/K$. In the transverse direction, $N_{\mathrm{max}}$ acts as truncation for the total energy of the 2D-HO basis states as: $N_{\mathrm{max}}\geq\sum\limits_i(2n_i+|m_i|+1)$. In addition, $N_{\mathrm{max}}$ implies the infrared (IR) and ultraviolet (UV) cutoffs in the transverse direction as $\lambda_{\mathrm{IR}}\backsimeq b/\sqrt{N_{\mathrm{max}}}$ and $\lambda_{\mathrm{UV}}\backsimeq b\sqrt{N_{\mathrm{max}}}$ \cite{IR_UV_cutoff}, respectively. 

After diagonalizing the Hamiltonian matrix, we obtain the proton mass $M$  and the corresponding LFWF in momentum space expressed using the obtained eigen vector as 
\begin{align}
    \Psi^\Lambda_{\mathcal{N}}(\{p_i,\lambda_i\})=\sum\limits_{\{n_i,m_i\}}\psi_\mathcal{N}^\Lambda(\{\alpha_i\}) \prod\limits^{\mathcal{N}}_{i=1}\phi_{n_i,m_i}(\Vec{p}^\perp_i,b). 
\end{align}
Here $\alpha_i$ are the set of quantum numbers $k_i,n_i,m_i,\lambda_i$ of each Fock-particle, and $\psi^\Lambda_{\mathcal{N}=3}(\{\alpha_i\})$ and $\psi^\Lambda_{\mathcal{N}=4}(\{\alpha_i\})$ are the components of eigenvectors associated with the Fock-sectors $|qqq\rangle$ and $|qqqg\rangle$, respectively. 

All the calculations in this work are performed with cutoffs $\{N_\mathrm{max},K\}=\{9,16.5\}$, and parameters are summarized in Table.~\ref{model parameters}, which were set by fitting the proton mass and its flavor FFs \cite{Siqi2022qqqg}. 
\begin{table}[h]
  \caption{The model parameters for cutoffs $\{N_{\text{max}},K\}=\{9,16.5\}$~\cite{Siqi2022qqqg}. All parameters are in unit of GeV except $g_s$. }
  \vspace{0.15cm}
  \label{model parameters}
  \centering
  \begin{tabular}{ccccccc}
    \hline\hline
         $m_u$ & $m_d$ & $m_g$ & $\kappa$ & ${m}_f$ & $b$ & $g_s$\\        
    \hline 
        0.31 & 0.25 & 0.5 & 0.54 & 1.80 & 0.7 & 2.2 \\       
    \hline\hline
  \end{tabular}
\end{table}

\section{Gluon transverse-momentum-dependent parton distributions \label{Sec3}}
The gluon TMDs at leading twist are given by parameterizing the following  gluon-gluon correlation function \cite{more_in_TMD}
\begin{align}
    &\Phi^{g[ij]}(x,\boldsymbol{k}_\perp;S)=\frac{1}{xP^+}\int\frac{dz^-}{2\pi}\frac{d^2\boldsymbol{z}_\perp}{(2\pi)^2}e^{ikz} \nonumber\\
    &\times\langle P;S|F^{+j}_a(0)\mathcal{W}_{+\infty,ab}(0;z)F^{+i}_b(z)|P;S\rangle\mid_{z^+=0^+}. 
    \label{correlator}
\end{align}
Here $a$ and $b$ are color indexes, $F^{+i}$ and $F^{+j}$ represent the gauge tensor with $i,j=1,2$ taking values from the transverse direction. $|P;S\rangle$ is the bound state of the proton with momentum $P$ and spin vector $S$. 
In this work, we only retain the zeroth-order expansion of the gauge link $\mathcal{W}\approx\mathbf{1}$, under which choice all time-reversal-odd (T-odd) TMDs would be zero \cite{gauge-link1,gauge-link2,gauge-link3,gauge-link4}. 

The twist-2 gluon TMDs are related to the correlator Eq.(\ref{correlator})  as \cite{more_in_TMD}, 
\begin{align}
    &\Phi^g(x,\boldsymbol{k}_\perp;S)=\delta^{ij}_\perp\Phi^{g[ij]}(x,\boldsymbol{k}_\perp;S)\nonumber\\
    &=f^g_1(x,\boldsymbol{k}_\perp^2)-\frac{\epsilon^{ij}_\perp k^i_\perp S^j_\perp}{M} f^{\perp g}_{1T}(x,\boldsymbol{k}_\perp^2), \\
    &\Tilde{\Phi}^g(x,\boldsymbol{k}_\perp;S)=i\epsilon^{ij}_\perp\Phi^{g[ij]}(x,\boldsymbol{k}_\perp;S)\nonumber\\
    &=S^3 g^g_{1L}(x,\boldsymbol{k}_\perp^2)+\frac{\boldsymbol{k}_\perp\cdot\boldsymbol{S}_\perp}{M}g^g_{1T}(x,\boldsymbol{k}_\perp^2),\\ 
    &\Phi^{g,ij}_T(x,\boldsymbol{k}_\perp;S)=-\boldsymbol{\hat{S}}\Phi^{g[ij]}(x,\boldsymbol{k}_\perp;S)\nonumber\\
    &=-\frac{\boldsymbol{\hat{S}} k^i_\perp k^j_\perp}{2M^2} h^{\perp g}_1 (x,\boldsymbol{k}_\perp^2)+\frac{S^3 \boldsymbol{\hat{S}} k^i_\perp \epsilon^{jk}_\perp k^k_\perp}{2M^2} h^{\perp g}_{1L} (x,\boldsymbol{k}_\perp^2)\nonumber\\
    &+\frac{\boldsymbol{\hat{S}} k^i_\perp \epsilon^{jk}_\perp S^k_\perp}{2M} \left(h^g_{1T}(x,\boldsymbol{k}_\perp^2)+\frac{\boldsymbol{k}_\perp^2}{2M^2} h^{\perp g}_{1T}(x,\boldsymbol{k}_\perp^2)\right)\nonumber\\
    &+\frac{\boldsymbol{\hat{S}} k^i_\perp \epsilon^{jk}_\perp (2k^k_{\perp}\boldsymbol{k}_\perp \cdot \boldsymbol{S}_\perp - S^k_\perp \boldsymbol{k}_\perp^2)}{4M^3}h^{\perp g}_{1T}(x,\boldsymbol{k}_\perp^2). 
\end{align}
Here $\boldsymbol{\hat{S}}$ is the symmetry operator defined as 
\begin{align}
    \boldsymbol{\hat{S}}O^{ij}=\frac{1}{2}(O^{ij}+O^{ji}-\delta^{ij}_\perp O^{mm})
\end{align}
for a generic tensor $O^{ij}$.  

The TMDs $f^{\perp g}_{1T}(x,\boldsymbol{k}_\perp^2),\ h^{\perp g}_{1L} (x,\boldsymbol{k}_\perp^2),\ h^g_{1T}(x,\boldsymbol{k}_\perp^2)$, and$\\h^{\perp g}_{1T}(x,\boldsymbol{k}_\perp^2)$ are T-odd so that they all vanish due to our choice of gauge link \cite{gauge-link1,gauge-link2,gauge-link3,gauge-link4}, The remaining four time-reversal-even (T-even)  leading-twist TMDs are expressed in terms of the LFWF in the $|qqqg\rangle$ Fock sector,  $\Psi^\Lambda_{\lambda_g,\lambda_{u1},\lambda_{u2},\lambda_d} (\{p_i\}) = \Psi^\Lambda_{\mathcal{N} = 4} (\{p_i,\lambda_i\})$, 
 
\begin{align}
    &f_1^g(x,\boldsymbol{k}_\perp^2)=\int [D] \sum\limits_{\lambda_{u_1}\lambda_{u_2}\lambda_{d}}\nonumber\\ &\times(\Psi_{+\lambda_{u_1}\lambda_{u_2}\lambda_{d}}^{+*} \Psi_{+\lambda_{u_1}\lambda_{u_2}\lambda_{d}}^+ +\Psi_{-\lambda_{u_1}\lambda_{u_2}\lambda_{d}}^{+*} \Psi_{-\lambda_{u_1}\lambda_{u_2}\lambda_{d}}^+), 
\end{align}

\begin{align}
    &g_{1L}^g(x,\boldsymbol{k}_\perp^2)=\int [D] \sum\limits_{\lambda_{u_1}\lambda_{u_2}\lambda_{d}}\nonumber\\ &\times(\Psi_{+\lambda_{u_1}\lambda_{u_2}\lambda_{d}}^{+*} \Psi_{+\lambda_{u_1}\lambda_{u_2}\lambda_{d}}^+ -\Psi_{-\lambda_{u_1}\lambda_{u_2}\lambda_{d}}^{+*} \Psi_{-\lambda_{u_1}\lambda_{u_2}\lambda_{d}}^+), 
\end{align}

\begin{align}
    &g_{1T}^g(x,\boldsymbol{k}_\perp^2)=\frac{2M}{(k^1_\perp)^2+(k^2_\perp)^2}\int [D] \sum\limits_{\lambda_{u_1}\lambda_{u_2}\lambda_{d}}\nonumber\\ &\times\mathcal{R}e \left[(k^1_\perp+ik^2_\perp)\Psi_{+\lambda_{u_1}\lambda_{u_2}\lambda_{d}}^{+*}\Psi_{+\lambda_{u_1}\lambda_{u_2}\lambda_{d}}^{-} \right], 
\end{align}

\begin{align}
    &h_1^{\perp g}(x,\boldsymbol{k}_\perp^2)=\frac{2M}{(k^1_\perp)^2-(k^2_\perp)^2}\int [D]\sum\limits_{\lambda_{u_1}\lambda_{u_2}\lambda_{d}}\nonumber\\&\times (\Psi_{+\lambda_{u_1}\lambda_{u_2}\lambda_{d}}^{+*}\Psi_{-\lambda_{u_1}\lambda_{u_2}\lambda_{d}}^{+}+\Psi_{-\lambda_{u_1}\lambda_{u_2}\lambda_{d}}^{+*}\Psi_{+\lambda_{u_1}\lambda_{u_2}\lambda_{d}}^{+}). 
\end{align}
Here we define a simplifying notation
\begin{align}
    &[D]\equiv\frac{1}{[2(2\pi)^3]^3}\prod\limits^4_{i=1} (dx_i d^2 \boldsymbol{k}_\perp)\nonumber \\&\times\delta(1-\sum\limits^4_{i=1}x_i) \delta^2(\sum\limits^4_{i=1} \boldsymbol{k_\perp}_i)
    \delta(x-x_g)\delta^2(\boldsymbol{k_{\perp}}-\boldsymbol{k_{\perp}}_g), 
\end{align}
and omit the argument $(\{p_i\})$ of each LFWF. 

\begin{figure*}[h]
    \centering
    \includegraphics[width=0.45\textwidth]{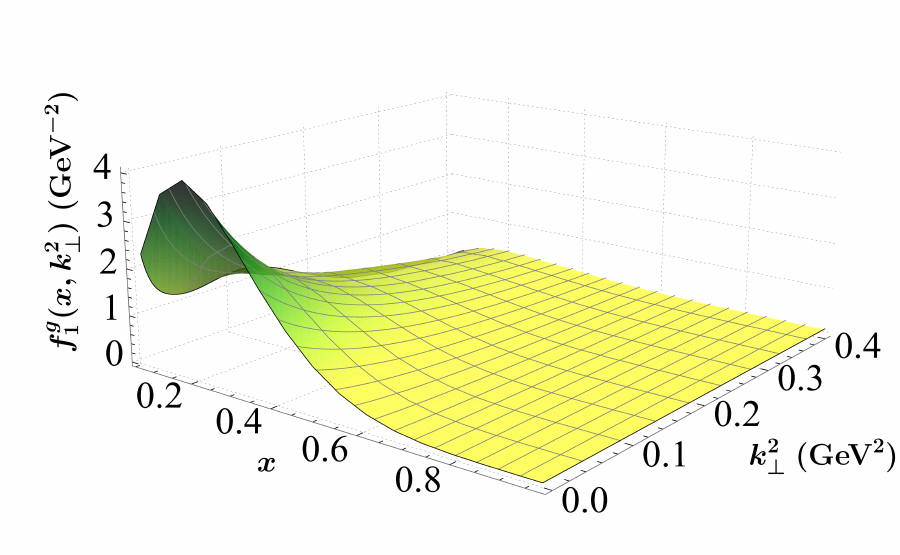}
    \includegraphics[width=0.45\textwidth]{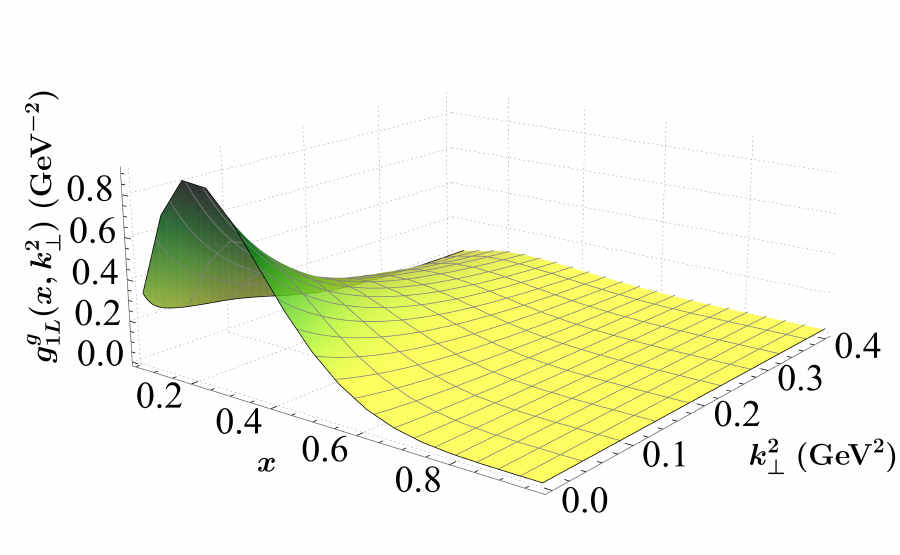}
    \includegraphics[width=0.45\textwidth]{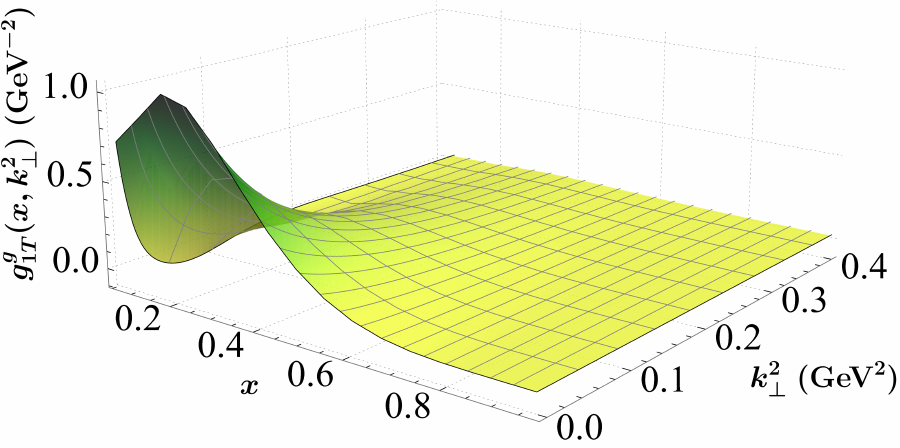}
    \includegraphics[width=0.45\textwidth]{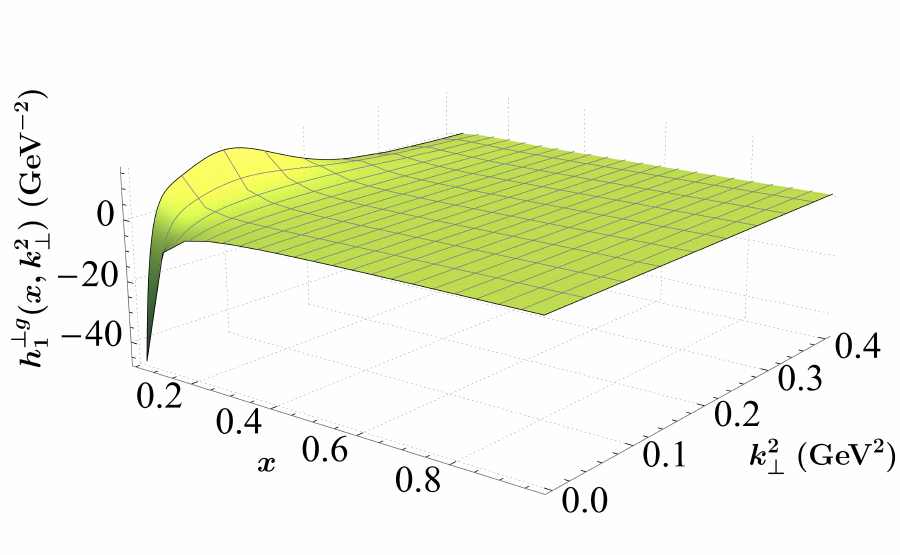}    
    \caption{\label{3D Plots}Four T-even gluon TMDs as functions of $x$ and $\boldsymbol{k}_\perp^2$. Left upper panel: the unpolarized TMD $f_1^g(x,\boldsymbol{k}_\perp^2)$; right upper panel: the helicity TMD $g_{1L}^g(x,\boldsymbol{k}_\perp^2)$; left lower panel: the worm-gear TMD $g_{1T}^g(x,\boldsymbol{k}_\perp^2)$; right lower panel: the Boer-Mulders TMD $h_1^{\perp g}(x,\boldsymbol{k}_\perp^2)$.}
\end{figure*}

We assemble our results of those TMDs in Fig.~\ref{3D Plots}. Our numerical results are oscillatory in the small-$x$ region, originating purely from the behaviour of the limited number of 2D-HO basis functions. Despite this numerical artifact, the BLFQ calculations for $f_1^g(x,\boldsymbol{k}_\perp^2)$, $g_{1L}^g(x,\boldsymbol{k}_\perp^2)$, and $g_{1T}^g(x,\boldsymbol{k}_\perp^2)$ are positive and have peaks in the small-$x$, small-$k_\perp^2$ region. The zero-crossing behaviour of $h_1^{\perp g}(x,\boldsymbol{k}_\perp^2)$ is persistent with different cutoffs and thus is a physical prediction of the BLFQ calculations. In the light-cone spectator model \cite{Chakrabarti_2023} and the AdS/QCD calculations \cite{Sumrule}, all four T-even gluon TMDs are positive, while in the spectator model \cite{Bacchetta2020}, $f_1^g(x,\boldsymbol{k}_\perp^2)$, $g_{1L}^g(x,\boldsymbol{k}_\perp^2)$, and $h_1^{\perp g}(x,\boldsymbol{k}_\perp^2)$ are positive but $g_{1T}^g(x,\boldsymbol{k}_\perp^2)$ is negative. We find that $f_1^g(x,\boldsymbol{k}_\perp^2)$ and $g_{1L}^g(x,\boldsymbol{k}_\perp^2)$ are always positive for all those calculations. Due to the discrete nature of the BLFQ framework for the longitudinal momentum fraction, we can only constrain this conclusion to the large-$x$ ($\gtrapprox 0.1$) region currently available to the BLFQ calculations. This behaviour is consistent with the recent extraction of the gluon PDFs $f_1^g(x)$ and $g_{1L}^g(x)$ \cite{PhysRevD.103.036007,PhysRevD.105.114017,Ball:2017otu,NNPDF:2017mvq,Harland-Lang:2014zoa,adamiak2023global,Nocera:2014gqa} and the lattice calculations of their moments \cite{Alexandrou:2020sml}. On the other hand, currently, there is insufficient experimental data or theoretical argument to constrain the distribution $g_{1T}^g(x,\boldsymbol{k}_\perp^2)$ and $h_1^{\perp g}(x,\boldsymbol{k}_\perp^2)$, and thus we simply present our results in Fig.~\ref{3D Plots} as the BLFQ predictions for these quantities.

\begin{figure*}[h]
    \centering
    \includegraphics[width=0.45\textwidth]{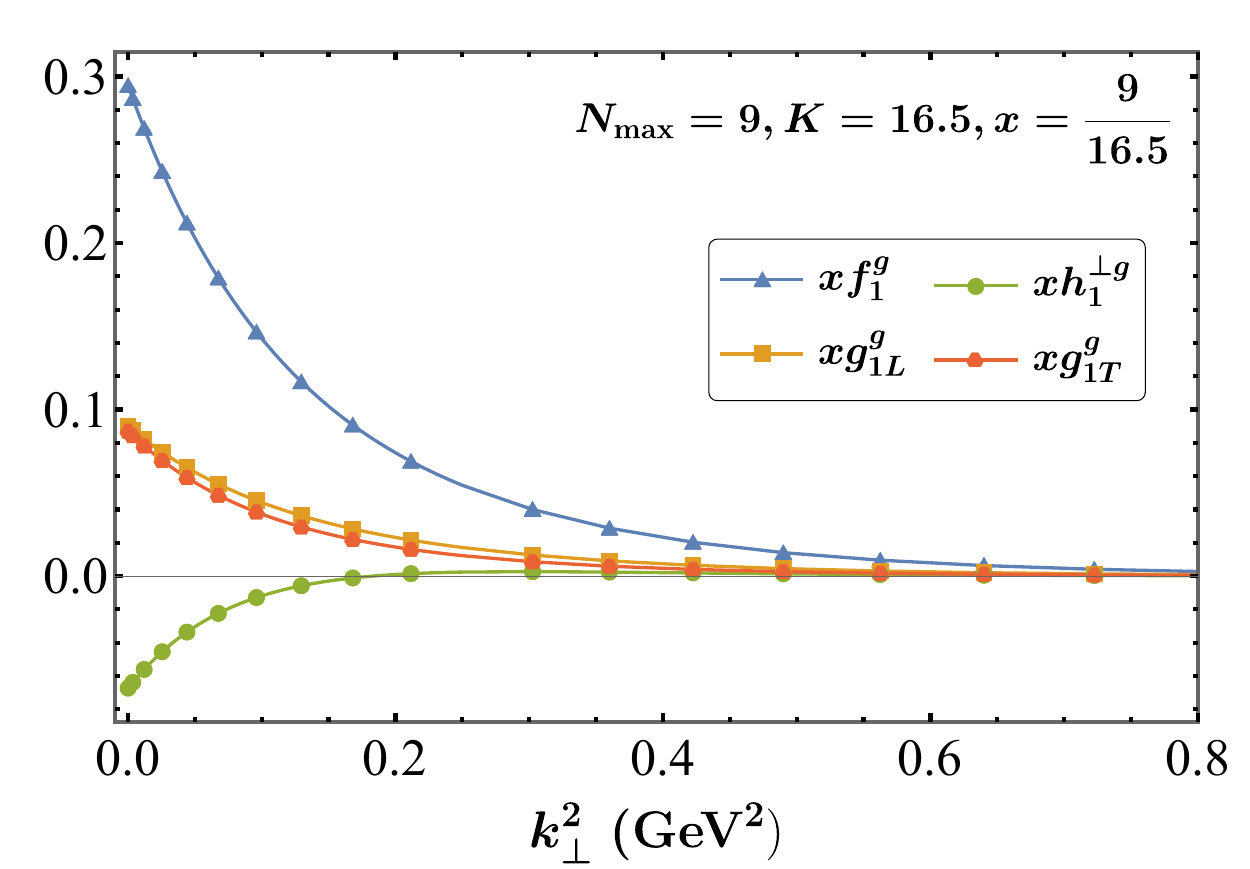}
    \includegraphics[width=0.45\textwidth]{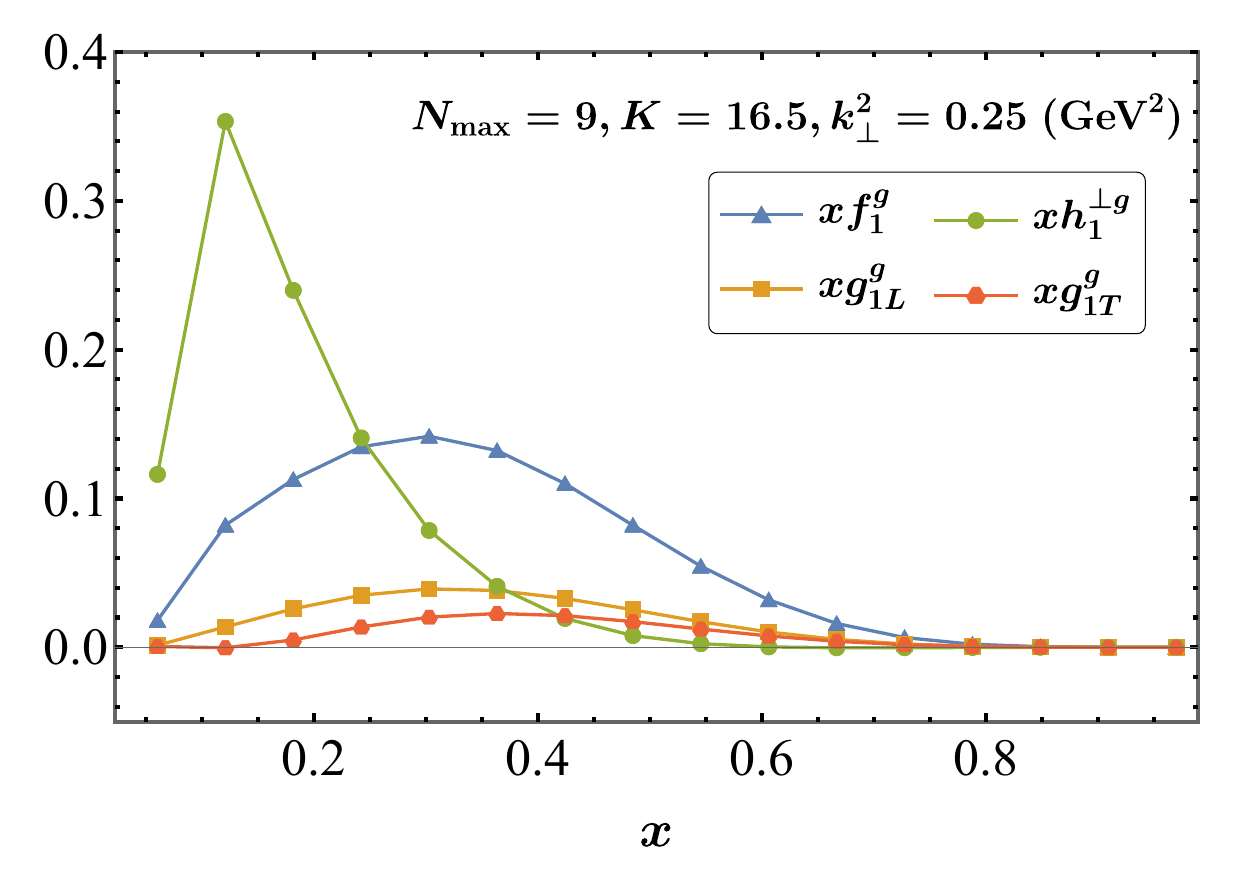}
    \caption{\label{Slices of TMD}Four T-even gluon TMDs in the transvers direction at $x=9/16.5$ (left panel), and in the longitudinal direction at $\boldsymbol{k}_\perp^2=0.25\ \mathrm{(GeV)^2}$ (right panel). }
\end{figure*}

\begin{figure}[h]
    \centering
    \includegraphics[width=0.45\textwidth]{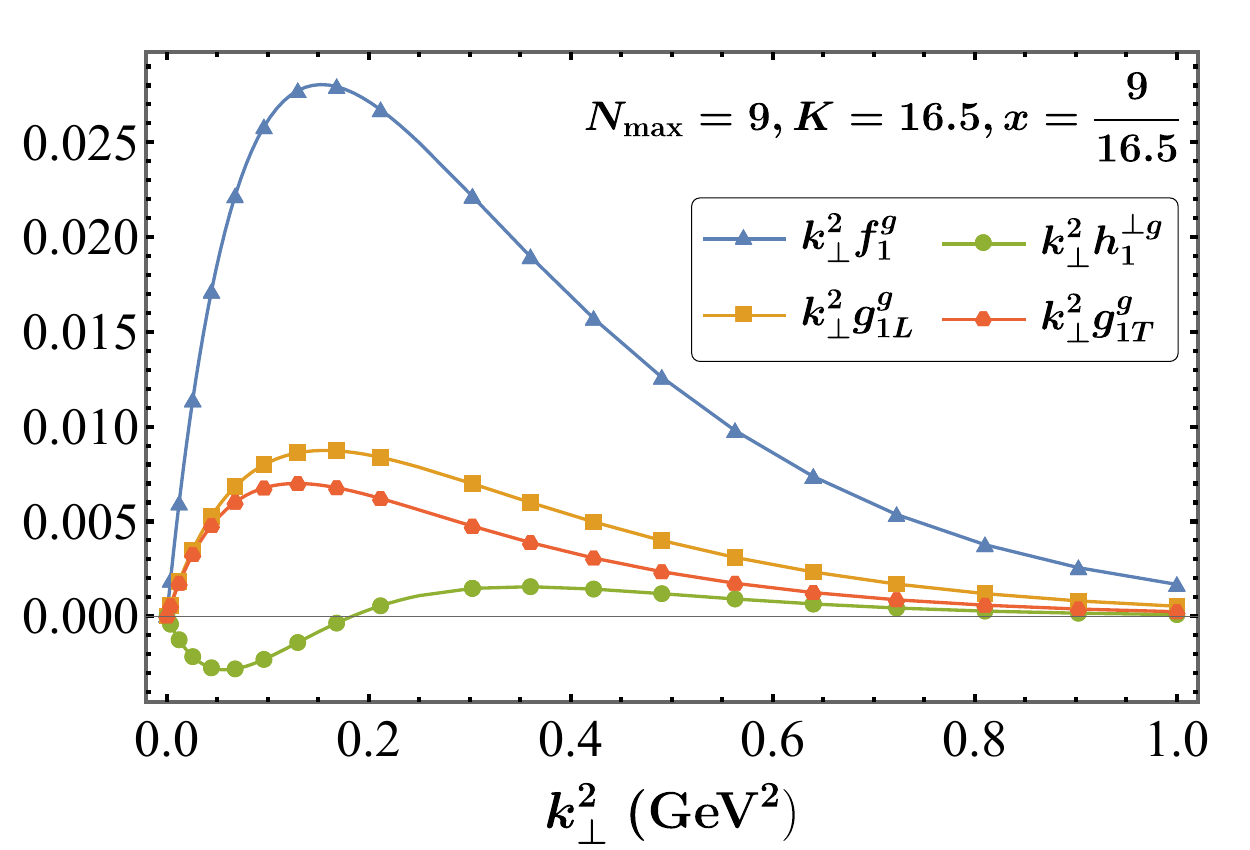}
    \caption{\label{small kperp2}Four T-even gluon TMDs multiplied by $\boldsymbol{k}_\perp^2$ in the transverse direction at $x=9/16.5$.}
\end{figure}

In Fig.~\ref{Slices of TMD}, we present our numerical results for the T-even gluon TMDs as functions of $\boldsymbol{k}_\perp^2$ at $x=9/16.5$ (left panel) and as functions of $x$ at $\boldsymbol{k}_\perp^2=0.25\ \mathrm{(GeV)^2}$ (right panel). In our numerical results, the absolute value of all four TMDs decreases monotonically as $\boldsymbol{k}_\perp^2$ increases. This feature arises since the S wave components dominate in the BLFQ calculations of all four TMDs. We also find that all four T-even TMDs tend to have very large values when $\boldsymbol{k}_\perp^2$ approaches zero. To study how fast they grow, we multiply all TMDs with $\boldsymbol{k}_\perp^2$ and the results are shown in Fig.~\ref{small kperp2}. We can see that as $\boldsymbol{k}^2_\perp$ approaches zero, products of $\boldsymbol{k}^2_\perp$ and TMDs also approaches zero. This shows that TMDs grow slower than $1/\boldsymbol{k}^2_\perp$ when $\boldsymbol{k}^2_\perp$ approaches zero.

\section{Discussions\label{Sec5}}

\subsection{Positivity bounds for gluon TMDs}

When all T-odd gluon TMDs vanish, the positivity bounds from Ref.  \cite{Cotogno:2017puy} reduce to the Mulders-Rodrigues inequalities \cite{Mulders:2000sh,Chakrabarti_2023,Sumrule}: 



\begin{align}
    \label{MR1}
    f_1^g(x,\boldsymbol{k}_\perp^2)\geq \sqrt{\left[g_{1L}^g(x,\boldsymbol{k}_\perp^2)\right]^2}, 
\end{align}
\begin{align}
    \label{MR2}
    f_1^g(x,\boldsymbol{k}_\perp^2)\geq\sqrt{\left[g_{1L}^g(x,\boldsymbol{k}_\perp^2)\right]^2+\left[\frac{|\boldsymbol{k_\perp}|}{M}g_{1T}^g(x,\boldsymbol{k}_\perp^2)\right]^2},
\end{align}    
\begin{align}    
    \label{MR3}
    f_1^g(x,\boldsymbol{k}_\perp^2)\geq\sqrt{\left[g_{1L}^g(x,\boldsymbol{k}_\perp^2)\right]^2+\left[\frac{|\boldsymbol{k_\perp}|^2}{2M^2}h_1^{\perp g}(x,\boldsymbol{k}_\perp^2)\right]^2}. 
\end{align}

We further search for the points where the left-hand side (LHS) quantities and the right-hand side (RHS) quantities are nearest to each other for all three inequalities. During the search, we set $\Delta|\boldsymbol{k}_\perp|=0.01\  (\mathrm{GeV})$ and $\Delta x=1/16.5$ for the increments in the transverse and longitudinal directions respectively. It is interesting to see that, contrary to the quark results in Ref. \cite{Hu:2022ctr}, LHS and RHS quantities of Eqs. (\ref{MR1}-\ref{MR3}) get constantly closer to each other as $\boldsymbol{k}_\perp^2$ grows and $x$ approaches 1.

\subsection{Transverse moments of the unpolarized gluon TMD}
\begin{figure*}
    \centering
    \includegraphics[width=0.45\textwidth]{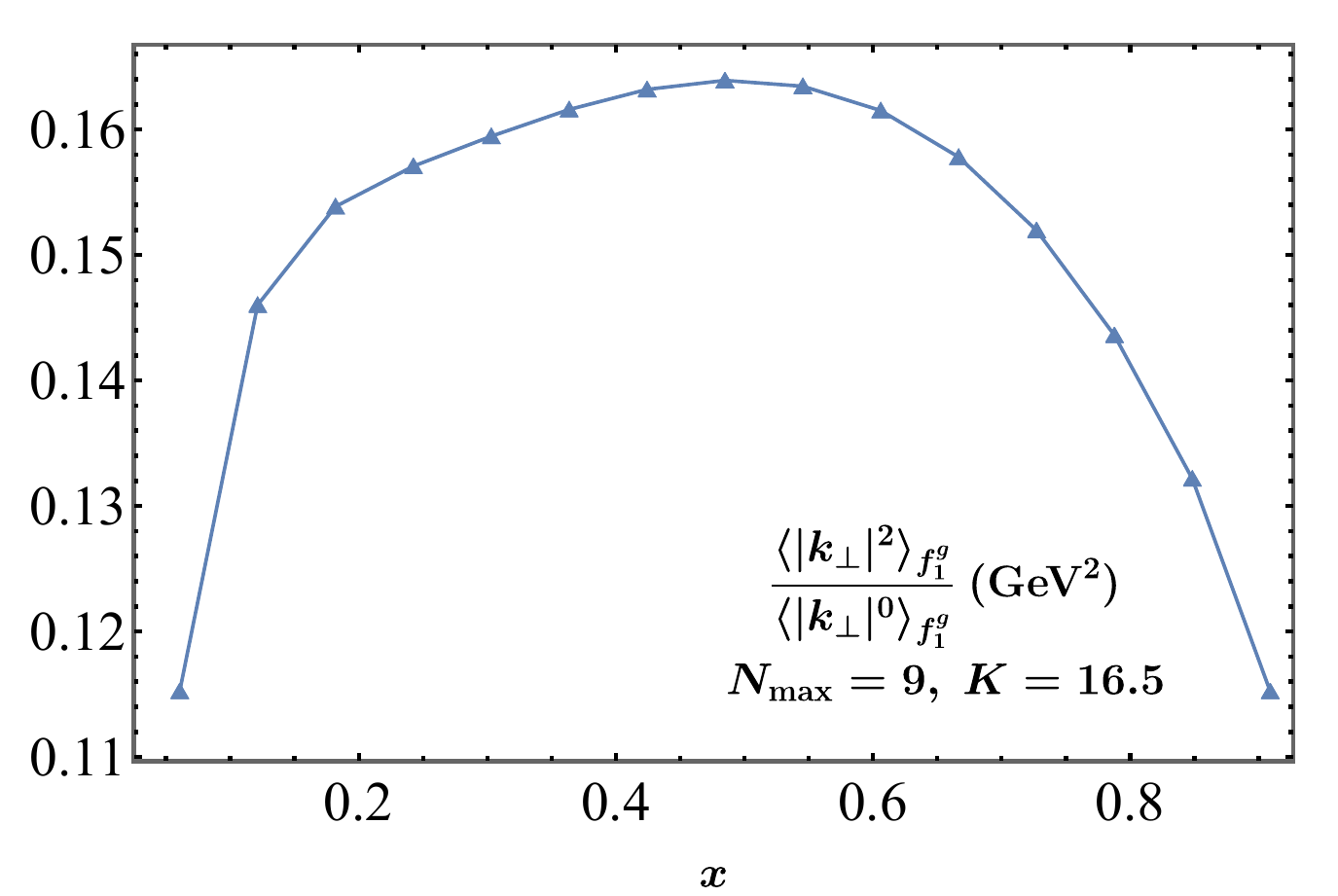}
    \includegraphics[width=0.45\textwidth]{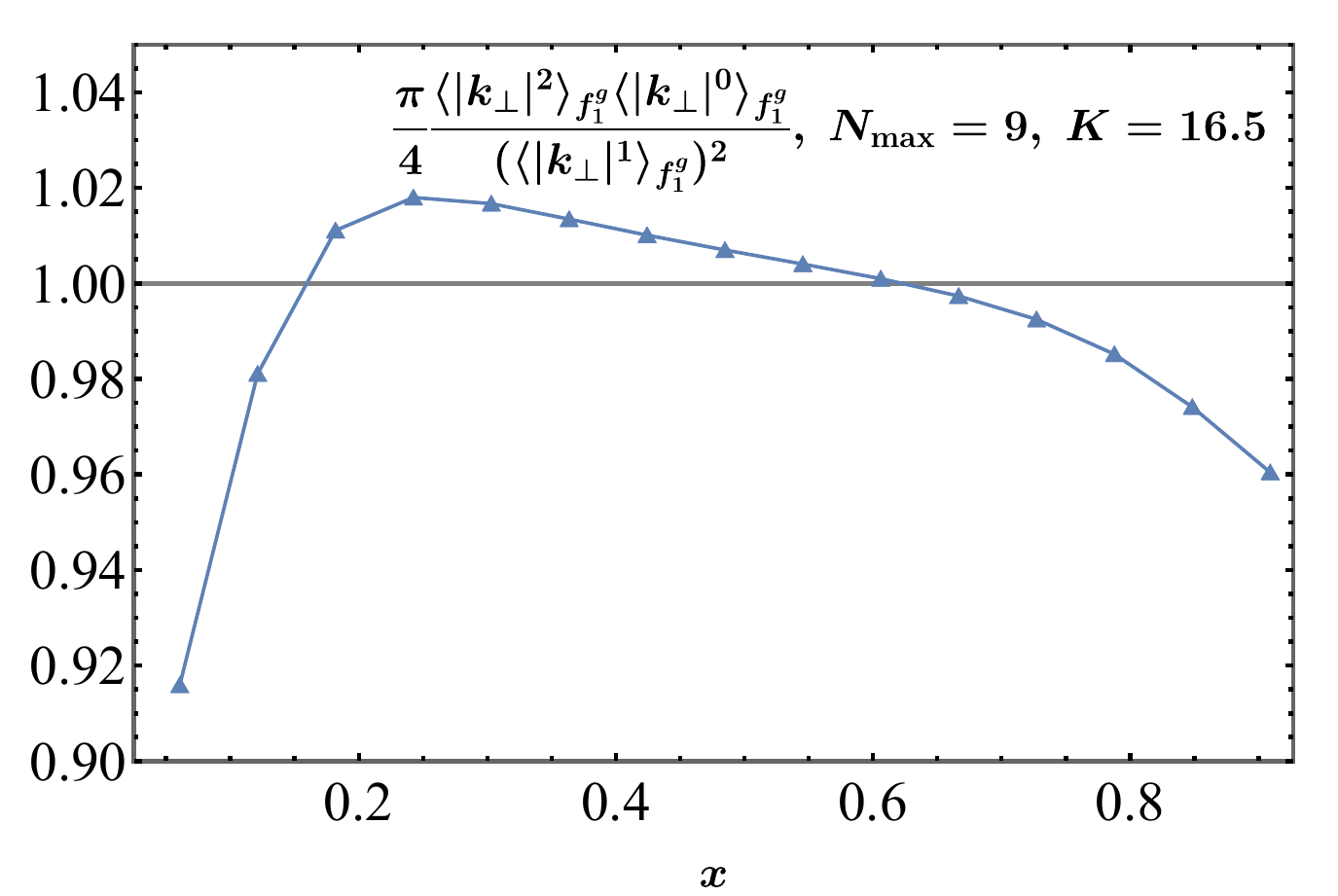}
    \caption{The left panel shows the BLFQ evaluations of the averaged transverse momentum squared of the unpolarized TMD, \ie, Eq. (\ref{Eq.average momentum 2}). The right panel presents the BLFQ evaluations of deviations of $f_1^g(x,\boldsymbol{k}_\perp^2)$ from a Gaussian distribution, i.e., Eq. (\ref{gaussian}).}
    \label{average momentum}
\end{figure*}

We define the $n$-th order transverse moment of the unpolarized gluon TMD as: 
\begin{align}
    \langle|\boldsymbol{k}_\perp|^n\rangle_{f_1^g}(x)=\int d^2\boldsymbol{k}_\perp |\boldsymbol{k}_\perp|^n f_1^g(x,\boldsymbol{k}_\perp^2).
    \label{Eq.average momentum}
\end{align}
From the second and the zeroth order moments, we can calculate the averaged transverse momentum squared by 
\begin{align}
   \frac{\langle|\boldsymbol{k}_\perp|^2\rangle_{f_1^g}}{\langle|\boldsymbol{k}_\perp|^0\rangle_{f_1^g}}. 
   \label{Eq.average momentum 2}
\end{align}
The BLFQ results for this averaged transverse momentum squared are shown in the left panel of Fig.~\ref{average momentum} as a function of the longitudinal momentum fraction. In other words, Fig.~\ref{average momentum} presents a correlation between the gluon's transverse and longitudinal degrees of freedom. The BLFQ calculations of Eq. (\ref{Eq.average momentum 2}) are relatively larger in the middle-$x$ region and decrease rapidly at large-$x$ and small-$x$ regions.

It is interesting to probe the compatibility of our BLFQ results with the Gaussian ansatz, \ie, whether we could model the BLFQ unpolarized gluon TMD with
\begin{align}
    f_1^g(x,\boldsymbol{k}_\perp^2)\approx a  \frac{\exp({-\frac{|\boldsymbol{k}_\perp|^2}{r}})}{\pi r}.
    \label{gaussian}
\end{align}
It is easy to show that under Gaussian ansatz,  $a=\langle|\boldsymbol{k}_\perp|^0\rangle_{f_1^g}$ and $r=\langle|\boldsymbol{k}_\perp|^2\rangle_{f_1^g}$. Thus, if the Gaussian ansatz holds, the following quantity
\begin{align}
    \frac{\langle|\boldsymbol{k}_\perp|^2\rangle_{f_1^g} \times \langle|\boldsymbol{k}_\perp|^0\rangle_{f_1^g}}{(\langle|\boldsymbol{k}_\perp|^1\rangle_{f_1^g})^2}\times\frac{\pi}{4}
    \label{gaussian ansatz}
\end{align}
should be unity at every $x$. The deviation of the value of Eq. (\ref{gaussian ansatz}) from 1 is a measure of the deviation of the unpolarized TMD from a Gaussian distribution. 
 
 In the right panel of Fig.~\ref{average momentum}, we show the results of Eq. (\ref{gaussian ansatz}) evaluated with the BLFQ results. We find that in the middle-$x$ region, our results are close to a Gaussian distribution and this measure is relatively stable. But at large-$x$ and small-$x$ regions, large deviations are evident. Since we work in a transverse 2D HO basis in which the transverse kinetic plus confinement portion of the Hamiltonian is diagonal, we can surmise that the Gaussian-like behavior in the middle-$x$ region reflects the dominance of kinetic plus confining physics in that region. On the other hand, the deviations in the small-$x$ and large-$x$ regions from simple Gaussian behavior signals a rise in the importance of gluon dynamics in these regions.

\subsection{Limiting behaviours of gluon TMDs}
As also evident from Fig.~\ref{average momentum}, the gluon TMDs are found to behave differently in the small-$x$ and large-$x$ regions than in the middle-$x$ region. Despite those currently little-known complicated correlations between the transverse and longitudinal degrees of freedom, there also exists theoretical requirements on the limiting behaviours of the transverse moments of the gluon TMDs. 

\begin{figure*}[h]
    \centering
    \includegraphics[width=0.45\textwidth]{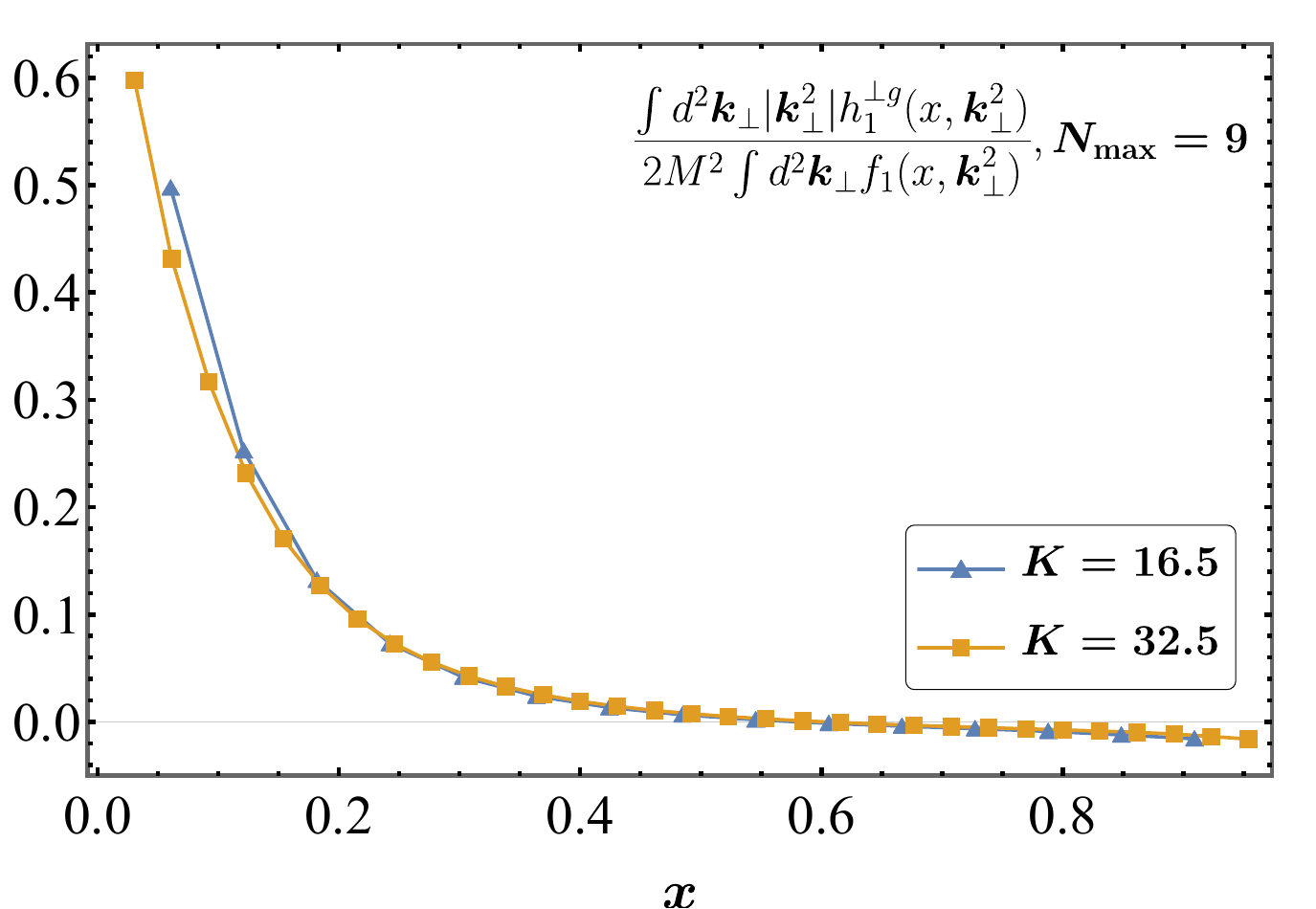}
    \includegraphics[width=0.45\textwidth]{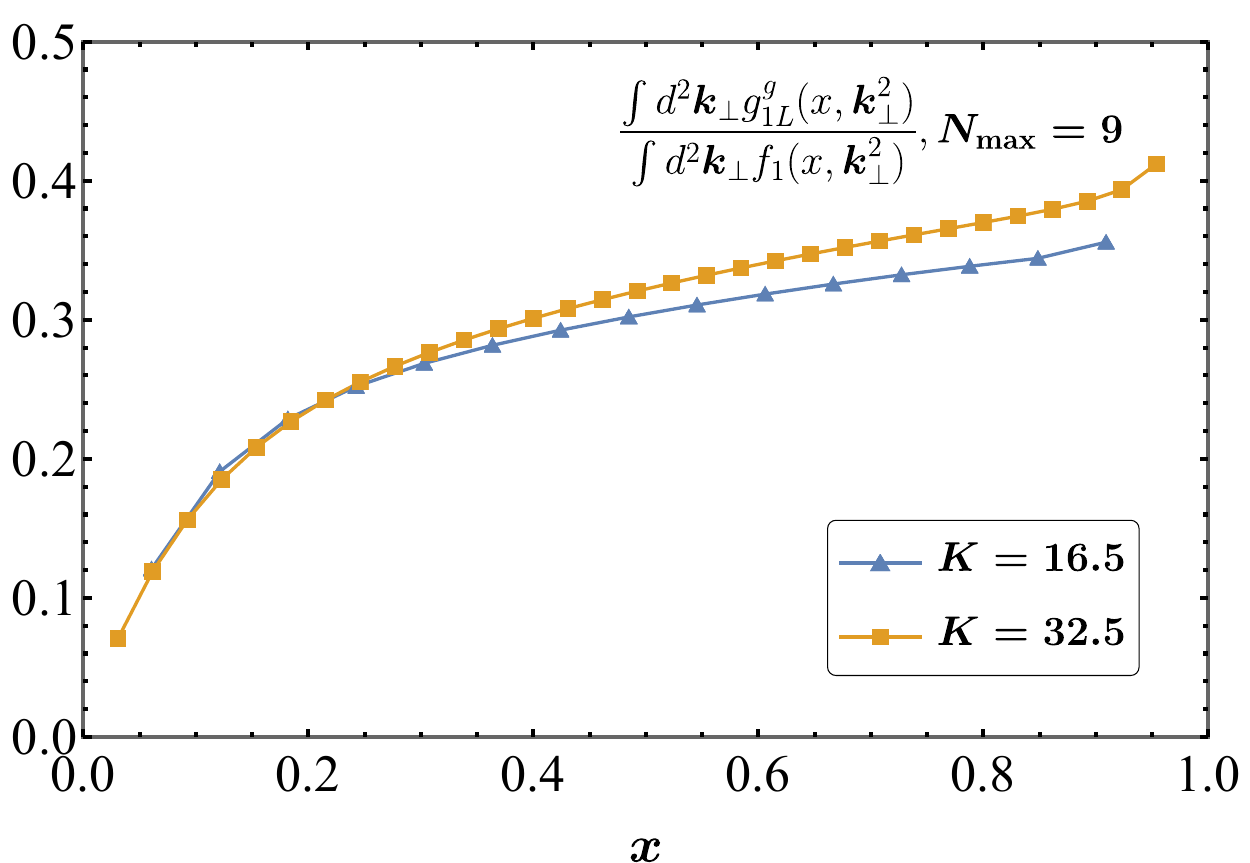}
    \caption{The left panel presents the BLFQ calculations of the ratio in Eq. (\ref{Eq.h1 Small-x}) from Ref. \cite{Smallx_Boer_2016}. The right panel displays the BLFQ calculations of the ratio in Eqs. (\ref{Eq:helicity asymmetry1}, \ref{Eq:helicity asymmetry2}) from Refs. \cite{Brodsky:1989db,Brodsky:1994kg}. We plot both ratios with $K=16.5$ and $K=32.5$ to investigate their trends as $x$ approaches the end points. }
    \label{fig:h1 Small-x}
\end{figure*}

As an example, Ref. \cite{Smallx_Boer_2016} shows that:
\begin{align}
    \lim_{x\rightarrow 0}\frac{\int d^2\boldsymbol{k}_\perp|\boldsymbol{k}_\perp^2| h_1^{\perp g}(x,\boldsymbol{k}_\perp^2)}{2M^2\int d^2\boldsymbol{k}_\perp f_1^g(x,\boldsymbol{k}_\perp^2)}=1.
    \label{Eq.h1 Small-x}
\end{align}
We show the BLFQ calculations of the ratio from Eq. (\ref{Eq.h1 Small-x}) as a function of $x$ in the left panel of Fig.~\ref{fig:h1 Small-x}. It is evident that after a somewhat slow raise from $x=1$ to $x \approx 0.3$, this ratio rapidly increases in the small-$x$ region. This is a promising sign that the BLFQ calculations satisfy the limit Eq. (\ref{Eq.h1 Small-x}). But, due to the discrete longitudinal basis and the finite truncation $ K $, it is difficult to approach much closer to the $ x\rightarrow 0 $ limit with currently available computational resources. Instead, as a test, we perform the BLFQ calculations with the same parameters as those listed in Table.~\ref{model parameters} but double the longitudinal cutoffs $ \{N_\mathrm{max}, K\} = \{9,32.5\} $. The ratio from Eq.~(\ref{Eq.h1 Small-x}) calculated with those new results is also plotted in the left panel of Fig.~\ref{fig:h1 Small-x} for comparison. One can easily observe the extension of the trend to approach 1, supporting our impression that BLFQ results for the gluon are consistent with the limiting behaviour in Eq.~(\ref{Eq.h1 Small-x}).

In Refs. \cite{Brodsky:1989db,Brodsky:1994kg}, the authors derived two other model-independent limiting requirements for the gluon TMDs from QCD:
\begin{align}
    &\lim_{x\rightarrow 0} \frac{\int d^2\boldsymbol{k}_\perp g_{1L}^g(x,\boldsymbol{k}_\perp^2)}{\int d^2\boldsymbol{k}_\perp f_1^g(x,\boldsymbol{k}_\perp^2)}=0, \label{Eq:helicity asymmetry1}\\
    &\lim_{x\rightarrow 1} \frac{\int d^2\boldsymbol{k}_\perp g_{1L}^g(x,\boldsymbol{k}_\perp^2)}{\int d^2\boldsymbol{k}_\perp f_1^g(x,\boldsymbol{k}_\perp^2)}=1. 
    \label{Eq:helicity asymmetry2}
\end{align}
We plot the $K=16.5$ and $K=32.5$ calculations of the ratios from Eqs.~(\ref{Eq:helicity asymmetry1}) and (\ref{Eq:helicity asymmetry2}) both with $N_\mathrm{max}=9$ and parameters listed in Table.~\ref{model parameters} in the right panel of Fig.~\ref{fig:h1 Small-x}. One again observes only  mild changes in the middle-$x$ region and relatively rapid increase (decrease) while the longitudinal fraction approaches 1 (0), serving as promising indications that the BLFQ calculations satisfy Eqs.~(\ref{Eq:helicity asymmetry1}) and (\ref{Eq:helicity asymmetry2}).


\subsection{Density distributions}

With the gluon TMDs, we now investigate the density distributions of the gluon inside the proton in three-dimensional momentum space. With the presented four T-even TMDs at the leading twist, there are four distributions connected with different polarizations of the gluon and the proton \cite{Bacchetta2020,Chakrabarti_2023}. The unpolarized density 
\begin{align}
    x\rho(x,k_x,k_y)=xf_1^g(x,\boldsymbol{k}_\perp^2), 
    \label{U density equation}
\end{align} 
is the probability density of finding an unpolarized gluon in an unpolarized nucleon at different $x$ and $\boldsymbol{k}_\perp$. The helicity density 
\begin{align}
    x\rho^{\circlearrowleft/+}(x,k_x,k_y)=xf_1^g(x,\boldsymbol{k}_\perp^2)+xg_{1L}^g(x,\boldsymbol{k}_\perp^2),
    \label{H density equation}
\end{align} 
gives the probability density of finding a circularly polarized gluon in a longitudinally polarized nucleon at different $x$ and $\boldsymbol{k}_\perp$. The Boer-Mulders density describes the probability density of finding a gluon linearly polarized along the $x$-axis in an unpolarized nucleon at different $x$ and $\boldsymbol{k}_\perp$: 
\begin{align}
    x\rho^\leftrightarrow(x,k_x,k_y)=\frac{1}{2}[xf_1^g(x,\boldsymbol{k}_\perp^2)+\frac{k_x^2-k_y^2}{2M^2}xh_1^{\perp g}(x,\boldsymbol{k}_\perp^2)],
    \label{BM density equation}
\end{align}
The last one is, the worm-gear density 
\begin{align}
    x\rho^{\circlearrowleft/\rightarrow}(x,k_x,k_y)=xf_1^g(x,\boldsymbol{k}_\perp^2)-\frac{k_x}{M}xg_{1T}^g(x,\boldsymbol{k}_\perp^2),
    \label{WG density equation}
\end{align}
is the probability of finding a circularly polarized gluon in a nucleon polarized along the $x$-axis at different $x$ and $\boldsymbol{k}_\perp$. 

\begin{figure*}[h]
    \centering
    \includegraphics[width=0.4\textwidth]{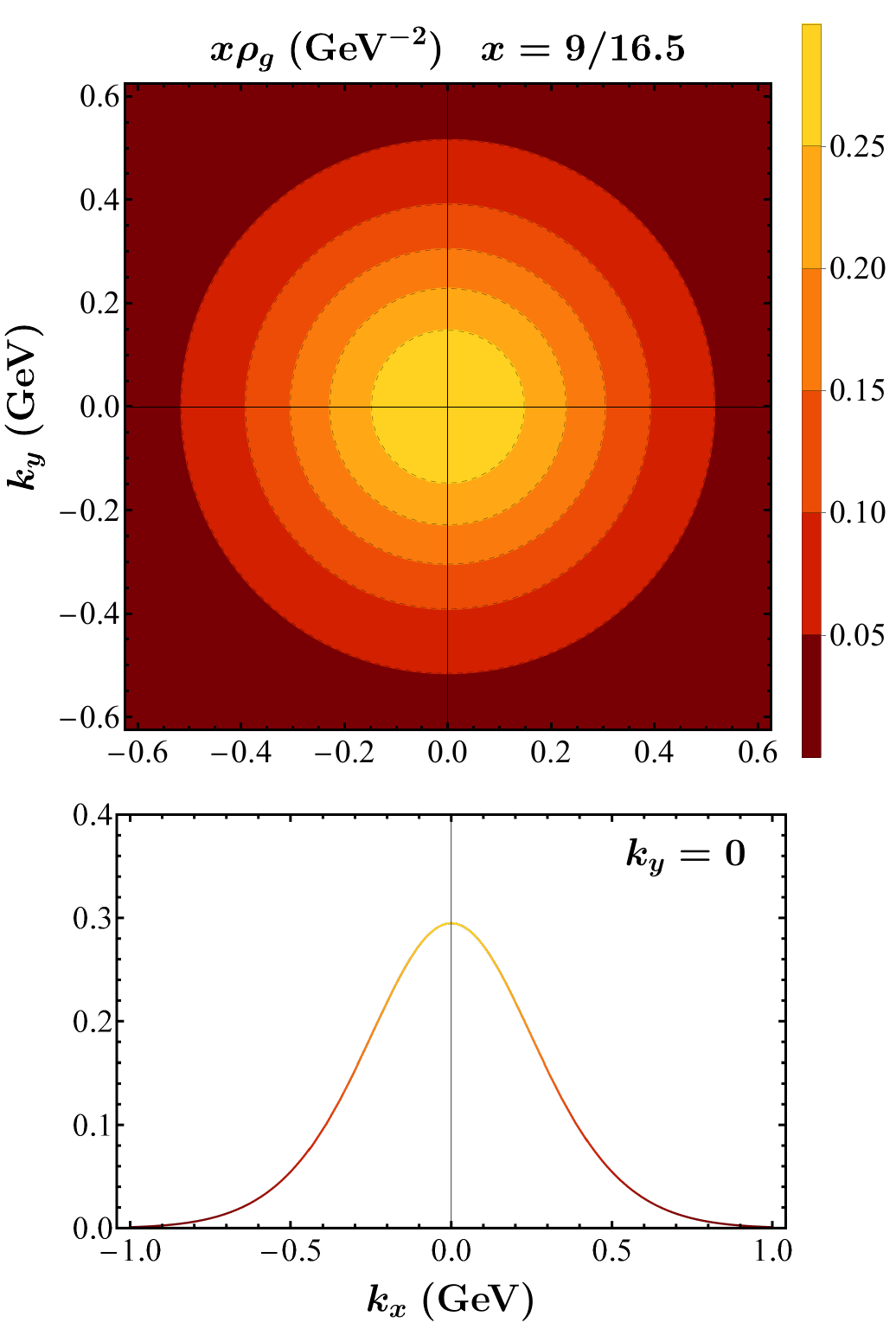}
    \includegraphics[width=0.4\textwidth]{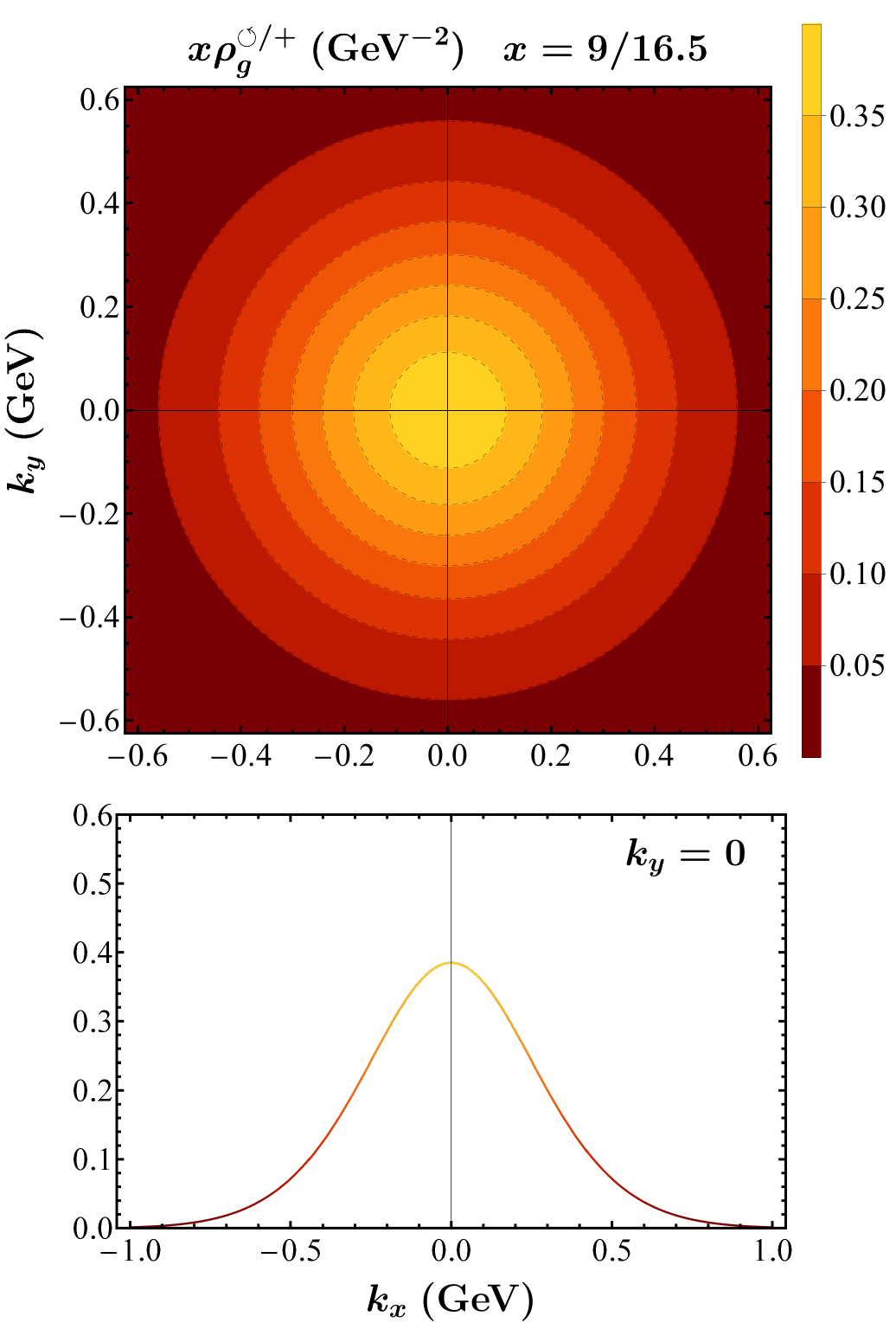}
    \includegraphics[width=0.4\textwidth]{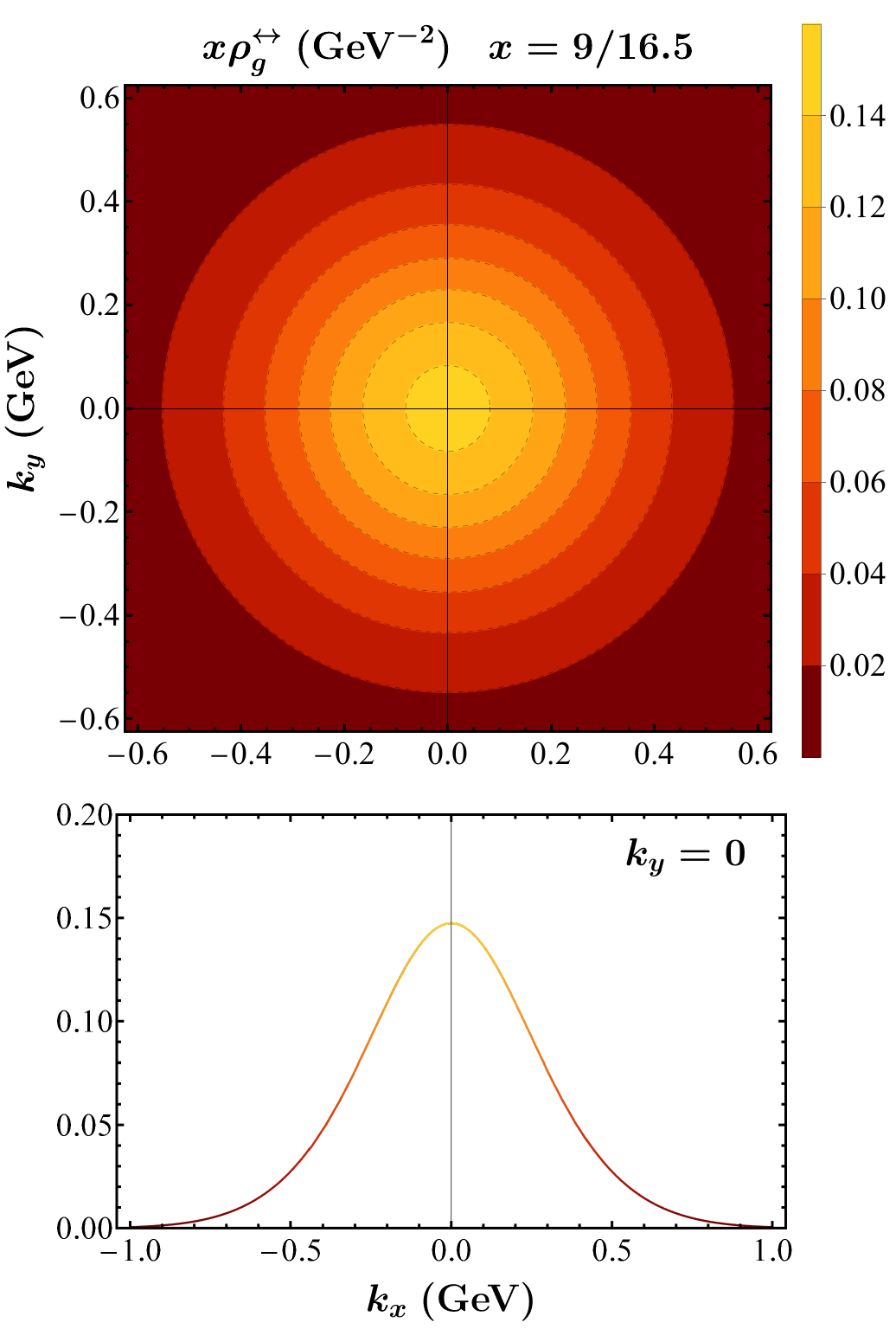}
    \includegraphics[width=0.4\textwidth]{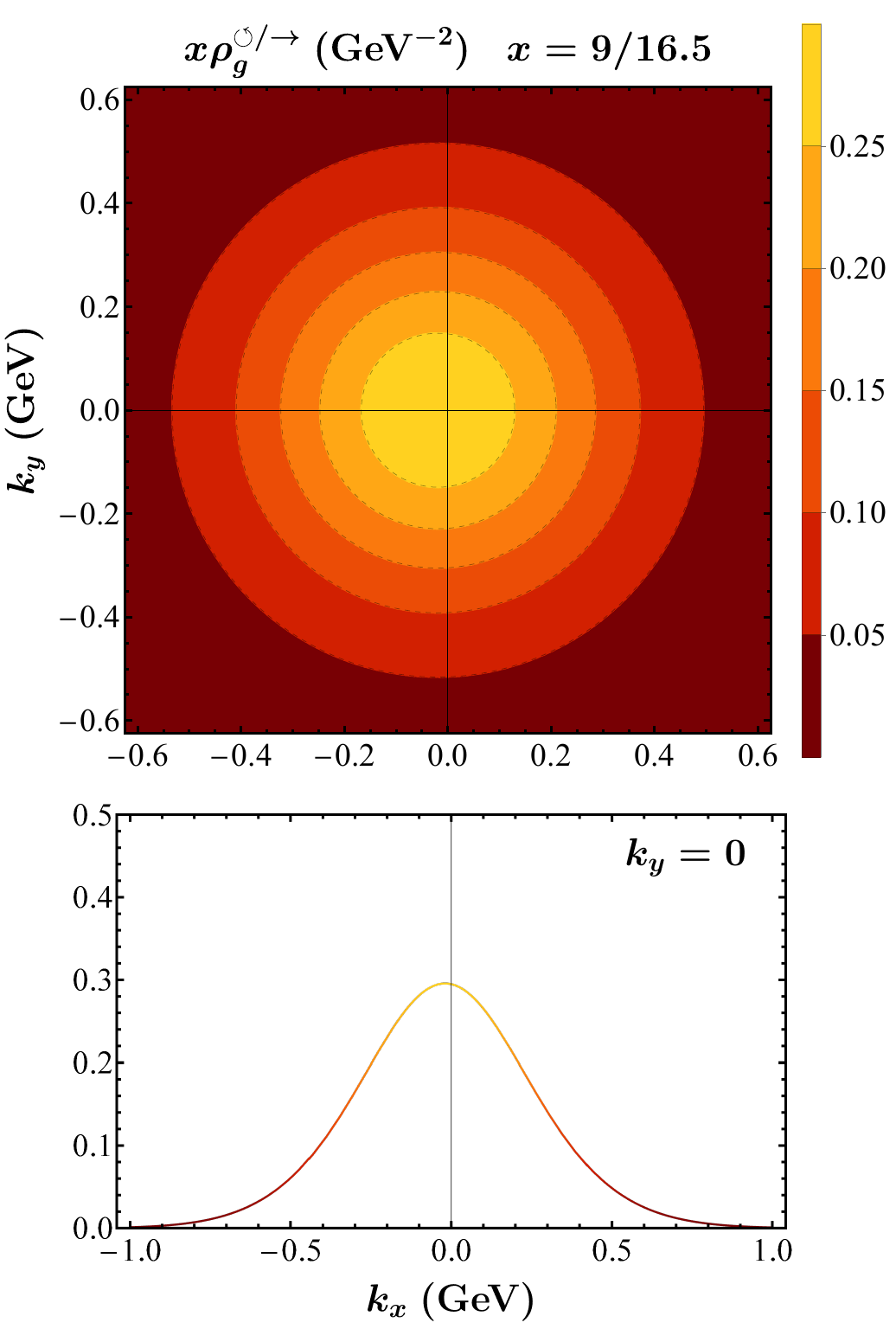}
    \caption{The gluon densities as functions of $\boldsymbol{k}_\perp$ and $x=\frac{9}{16.5}$, the 1D plot under each 2D density plot is the slice of the 2D plot above it at $k_y=0$. The left upper panel is the unpolarized gluon density, the right upper panel is the helicity density, and the left lower panel is the for ``Boer-Mulders" density, the right lower panel shows the ``worm-gear" density. }
    \label{density polt}
\end{figure*}    

All four densities as function of the transverse momenta $ k_x, k_y $ at fixed longitudinal momentum fraction $ x= 9/16.5 $ are plotted in Fig.~\ref{density polt}. The unpolarized and helicity densities (two panels in the upper row of Fig.~\ref{density polt}) are rotationally symmetric in the transverse plane, as can also be seen from Eqs. (\ref{U density equation}) and (\ref{H density equation}). On the other hand, Eq. (\ref{BM density equation}) indicates a $\cos2\varphi$ asymmetry for the Boer-Mulders density, which is not very obvious in the left lower panel of Fig.~\ref{density polt}. The reason for this behaviour is that the value of $ \boldsymbol{k}_\perp^2 h_1^{\perp g}(x,\boldsymbol{k}_\perp^2) $ is much smaller than that of $ f_1^g(x,\boldsymbol{k}_\perp^2) $, as can be seen from Figs.~\ref{Slices of TMD} and \ref{small kperp2}. Thus at the numerical level, the BLFQ results of the gluon Boer-Mulders density are totally dominated by the unpolarized distribution. Even though still milder than the results from \cite{Bacchetta2020,Chakrabarti_2023}, the $\cos\varphi$ modulation of the worm-gear density in the right lower panel of Fig.~\ref{density polt} is more evident than that of the Boer-Mulders density. This is because in Eq. (\ref{WG density equation}), $ g_{1T}^g(x,\boldsymbol{k}_\perp^2) $ is multiplied only by $ \sqrt{\boldsymbol{k}_\perp^2} $.

Another significant difference of the BLFQ results from other calculations for the gluon density (Figs.~5, 6 of Ref. \cite{Bacchetta2020} and Fig.~6 of Ref. \cite{Chakrabarti_2023}) is that the BLFQ gluon results in Fig.~\ref{density polt} do not decrease to zero at zero transverse momenta. This is expected since in Sec. 3, we find that S wave components dominate the BLFQ calculations of all four TMDs.

\section{Summary\label{Sec6}}
In this paper, we calculate the T-even gluon TMDs at the leading twist within the BLFQ framework. Our results are consistent with the currently available theoretical calculations and the experimental extractions. However, unlike the results obtained in Refs. \cite{Bacchetta2020,Chakrabarti_2023,Sumrule}, the BLFQ gluon TMDs are dominated by the S wave components, leading to the monotonical behaviours in the transverse direction. Further, due to the limiting behaviours in the transverse direction of all four TMDs calculated within the BLFQ framework, all polarized densities available with T-even TMDs do not exhibit large asymmetry. It remains an open question whether the contributions from higher Fock-sectors in BLFQ introduce significant asymmetry in the gluon polarized densities. Our future work will address this issue.

Our calculations of gluon TMDs do not support the model relation proposed in Ref. \cite{Sumrule} that relates the square of $f_1^g(x,\boldsymbol{k}_\perp^2)$ to the sum of the squares of other three T-even gluon TMDs, validating its model-dependent origin. On the other hand, restrictions of the gluon TMDs can be obtained from purely QCD considerations. Here, we investigate two of them, the first being the Mulders-Rodrigues inequalities \cite{Cotogno:2017puy,Mulders_2001} and the second being the limiting behaviours of moments of TMDs from Refs. \cite{Smallx_Boer_2016,Brodsky:1989db,Brodsky:1994kg}. T-even gluon TMDs obtained within the BLFQ framework fulfill all of them, strengthening the validity of the BLFQ calculations and our future plans to further investigate gluon TMD-related experimental results.

Due to the dependences on both the transverse and longitudinal directions, it is very convenient and interesting to investigate the correlations between those two degrees of freedom with the help of TMDs. Here we point out two such correlations, Eqs.~(\ref{Eq.average momentum 2}) and (\ref{gaussian}), which we investigate using the BLFQ results. We find relatively mild correlations in the middle-$x$ region and strong correlations in the small-$x$ and large-$x$ regions. We interpret these results as indicating the more complicated dynamics of the gauge boson at the extremes of longitudinal momentum fraction.

\section*{Acknowledgements}
We thank Zhimin Zhu and Jiatong Wu for many useful assistance. C. M. is supported by new faculty startup funding by the Institute of Modern Physics, Chinese Academy of Sciences, Grant No. E129952YR0. C. M. also thanks the Chinese Academy of Sciences Presidents International Fellowship Initiative for the support via Grants No. 2021PM0023. X. Z. is supported by new faculty startup funding by the Institute of Modern Physics, Chinese Academy of Sciences, by Key Research Program of Frontier Sciences, Chinese Academy of Sciences, Grant No. ZDBS-LY-7020, by the Natural Science Foundation of Gansu Province, China, Grant No. 20JR10RA067, by the Foundation for Key Talents of Gansu Province, by the Central Funds Guiding the Local Science and Technology Development of Gansu Province, Grant No. 22ZY1QA006, by Gansu International Collaboration and Talents Recruitment Base of Particle Physics (2023-2027), by International Partnership Program of the Chinese Academy of Sciences, Grant No. 016GJHZ2022103FN, by National Natural Science Foundation of China, Grant No. 12375143, by the Strategic Priority Research Program of the Chinese Academy of Sciences, Grant No. XDB34000000 and by National Key R\&D Program of China, Grant No. 2023YFA1606903. J. P. V. is supported by the Department of Energy under Grant No. DE-SC0023692. This research use resources of the National Energy Research Scientific Computing Center (NERSC), a U.S. Department of Energy Office of Science User Facility operated under Contract No. DE-AC02-05CH11231 using NERSC award NP-ERCAP0020944 . A portion of the computational resources were also provided by Gansu Computing Center.

\bibliography{sample}
\biboptions{sort&compress}
\bibliographystyle{elsarticle-num}

\end{document}